\documentclass[debug]{rmaa}

\usepackage{paralist}

\usepackage{psfrag,color}

\usepackage[latin1]{inputenc}

\title{The Radio Continuum Source Projected Near HR 8799} 

\author{
  L. F. Rodr\'iguez\altaffilmark{1,2},
  L. A. Zapata\altaffilmark{1} 
}
\altaffiltext{1}{Instituto de Radioastronom\'\i{}a y Astrof\'\i{}sica,
  UNAM, M\'exico.}

\altaffiltext{2}{Mesoamerican Center for Theoretical Physics, 
UNACH, 
M\'exico}

\shortauthor{Rodr\'iguez \& Zapata}
\shorttitle{Radio Continuum Source Near HR 8799}

\fulladdresses{
\item Luis F. Rodr\'iguez and Luis A. Zapata: Instituto de Radioastronom\'\i{}a y
  Astrof\'\i{}sica, Universidad Nacional Aut\'onoma de M\'exico,
  Apartado Postal 3--72, 58090 Morelia, Michoac\'an, M\'exico
  (l.rodriguez@irya.unam.mx).}

\listofauthors{L. F. Rodr\'iguez \& L. A. Zapata}
\indexauthor{Rodr\'iguez, L. F.}
\indexauthor{Zapata, L. A.}

\abstract{HR 8799 is an A5/F0 V star where exoplanets were first
directly imaged. Four exoplanets were found within $\simeq 2\rlap.{''}0$ from the star.
Here we report the VLA detection of a
faint (19.1$\pm$2.7 $\mu$Jy) radio continuum (3.0 GHz) source projected at $\simeq 2\rlap.{''}2$ from the star. 
The \sl a priori \rm probability of finding a background 
source with this flux density within a radius of $2\rlap.{''}2$ is only 0.0046.
However, the astrometry
made with the VLA and ALMA images, separated by 5.5 years, indicates no significant proper motions 
and rules out the association of the radio source with
the HR 8799 system and suggests it is a background millimeter galaxy
with dust emission in the millimeter and
partially thick synchrotron emission in the centimeter.
}

\resumen{HR 8799 es una estrella A5/F0 V donde se detectaron por primera vez exoplanetas
con la t\'ecnica de im\'agenes directas. Se encontraron cuatro exoplanetas dentro de $\simeq 2\rlap.{''}0$ de la estrella.
Aqu\'i reportamos la detecci\'on con el VLA de una
fuente d\'ebil de radio continuo (19.1$\pm$2.7 $\mu$Jy) a una frecuencia de 3.0 GHz proyectada a $\simeq 2\rlap.{''}2$ de la estrella. %Esta fuente fue detectada inicialmente en el submilim\'etrico con el
La probabilidad \sl a priori \rm de encontrar una fuente de fondo
con esta densidad de flujo dentro de un radio de $2\rlap.{''}2$ es solo 0.0046.
Sin embargo, la astrometr\'ia
realizada con las im\'agenes del VLA y ALMA, separadas por 5.5  a\~nos,
indica que no hay movimientos propios significativos
y descarta la asociaci\'on de la fuente de radio con
el sistema HR 8799.
y sugiere que se trata de una galaxia milim\'etrica de fondo
con emisi\'on de polvo en el milim\'etrico
y emisi\'on sincrotr\'on parcialmente \'opticamente gruesa en el centim\'etrico.
}

\addkeyword{Astrometry}
\addkeyword{proper motions}
\addkeyword{Radio continuum: general}
\addkeyword{stars: individual (HR 8799)}

\begin{document}
% Typeset article header 
\maketitle
\section{Introduction}
\label{sec:intro}
In excess of 100 exoplanets have been discovered with the technique of direct imaging, as listed
in the NASA Exoplanet Archive (Akeson et al. 2013).
HR 8799 is the star where exoplanets were first
directly imaged. Four exoplanets were found within $\sim$80 au ($\sim 2''$ at the distance of 40.85 pc)
in the plane of the sky from the star (Marois et al. 2008; 2010).
The observed proper motions of the exoplanets around the star confirmed the association (Close \& Males 2010).
In addition to the four exoplanets, HR 8799 exhibits a debris disk (Booth et al.
2016; Wilner et al. 2018), extending from  $\simeq 2''$ ($\simeq$80 au)
to $\simeq 7''$ ($\simeq$280 au) in radius, with a clear central cavity and detectable in the sub-millimeter (Faramaz et al. 2021).
In addition, these authors report the presence of an 880 $\mu$m point source associated with the star.

In this paper we present sensitive VLA observations of the HR 8799 region obtained with the purpose of searching for emission
from the star or from one of its four exoplanets. In section 2 we discuss the observations, while in section 3 we interpret
the data. Finally, our conclusions are presented in section 4.

\section{Observations}
\label{sec:observations}

\subsection{VLA}
The data of project 12B-188 was obtained from the archives of the 
Karl G. Jansky Very Large Array (VLA) of NRAO\footnote{The National 
Radio Astronomy Observatory is a facility of the National Science Foundation operated
under cooperative agreement by Associated Universities, Inc.}. These observations were made in the
highest angular resolution A configuration during six epochs between 2012 October 31 and 2012 November 26.
The observations were made in the S-band continuum (2-4 GHz), with 16 spectral windows of 128 MHz each. These spectral
windows were divided in 64 channels of 2 MHz individual width. The amplitude calibrator was J0137+3309 (3C48) 
and the gain calibrator was J2254+2445.
The data were calibrated in the standard manner using the version 5.6.2-3 of the CASA (Common Astronomy Software Applications;  McMullin et al. 2007) package of NRAO and
the pipeline provided for VLA\footnote{https://science.nrao.edu/facilities/vla/data-processing/pipeline} observations. The data of the six epochs were concatenated in a single file to increase the signal-to-noise ratio. The images 
were made using a robust weighting of 2 (Briggs 1995), to optimize the sensitivity at the expense of losing some angular resolution. 

\subsection{ALMA}

The sub-millimeter continuum data (at 340 GHz) were obtained from the ALMA data archive. 
The project was made under the program 2016.1.00907.S (PI: V. Faramaz). For this study, we only 
used the 12 m array observations carried out from 2018 May 13 to June 1.  The data were taken using baselines 
ranging from 15 to 314 m (18 to 392 k$\lambda$). The ALMA digital correlator was configured with four spectral windows 
(SPWs), each one 2 GHz wide. Three of these SPWs were used for the continuum, and one for the detection of the 
CO(3$-$2) molecular line at a rest frequency of 345.79598990 GHz. Bright quasars J2148$+$0657, J2253$+$1608, 
and J2253$+$1608 were used as flux, bandpass, and gain phase calibrators.   The total time on-source was 4.5 hrs.  
The raw data were calibrated, and then imaged using the Common Astronomy Software Applications (CASA) version 5.1.1.  
The digital correlator was set up with three spectral windows with a bandwidth of 2 GHz (divided by 128 channels resulting in a channel width of 15.625 MHz) and one 
spectral window with a bandwidth of 1.875 GHz (divided by 3840 channels resulting in a channel width of 488.281 kHz). The CO(3$-$2) line was the only spectral line 
excluded during the process of the continuum construction.  We obtained an image rms noise 
for the continuum at 0.8 mm of 10 $\mu$Jy beam$^{-1}$ at an angular resolution of $0\rlap.{''}88 \times 0\rlap.{''}76$; 14$^{\circ}$. 
This angular resolution is very similar to that obtained in the VLA image (see caption of Figure 1). We used a robust parameter equal to 0.5 in the TCLEAN task, an adequate compromise between angular resolution and sensitivity. 
The parameters of the sub-millimeter source detected are given in Table 1.
The ALMA observations were not configured to detect any polarized emission.
For both the VLA and the ALMA observations we fitted the data in the image plane with the task IMFIT of CASA. 
The results of such fittings are given in Table 1. 
Our derived value for ALMA, 207$\pm$10 $\mu$Jy is lower than the value of  316$\pm$20 $\mu$Jy obtained
by Faramaz et al. (2021). This difference is probably due to the fact that Faramaz et al. (2021) included in
their analysis data from
the Atacama Compact Array, recovering more extended flux.

\begin{table*}[!t]\centering
  \scriptsize
 \newcommand{\DS}{\hspace{1\tabcolsep}} %% Expanded Space between
  %% some cols
  \begin{changemargin}{-3.0cm}{-2cm}
    \caption{Parameters of the VLA and ALMA observations of the HR 8799 Region}
    \setlength{\tabnotewidth}{1.05\linewidth}
    \setlength{\tabcolsep}{1.2\tabcolsep} \tablecols{9}
    \begin{tabular}{l @{\DS} cccccccc}
      \toprule
      % & \multicolumn{9}{c}{Ionization Stage}\\
      %\cmidrule{3-9}
      % & \multicolumn{4}{c}{Log (Radial Average)}
      & Mean &  & Frequency/ & Deconvolved & Total Flux &\multicolumn{2}{c}{Position\tabnotemark{a}}\\
      % \cmidrule(l){7-8}
      Project & Epoch & Telescope & Bandwidth & Dimensions & Density  & 
      RA(J2000)\tabnotemark{b} \label{tab:par} & DEC(J2000)\tabnotemark{c} \label{tab:par}  \\
      \midrule
       12B-188& 2012.88 &  VLA  & 3/2 GHz &  
      0$\rlap.{''}$5$\pm$0$\rlap.{''}$2$\times$0$\rlap.{''}$2$\pm$0$\rlap.{''}$2;$+71^{\circ}$$\pm35$$^{\circ}$ & 19.1$ \pm$2.7 $\mu$Jy
      & 28$\rlap.^{s}$763$\pm$0$\rlap.^{s}$003 & 
      04$\rlap.{''}$71$\pm$0$\rlap.{''}$03 \\  
       2016.1.00907.S & 2018.38 &  ALMA  & 340/8 GHz  & 
       0$\rlap.{''}$5$\pm$0$\rlap.{''}$1$\times$0$\rlap.{''}$5$\pm$0\rlap.{''}1;$+160^{\circ}$$\pm$$91^{\circ}$ & 
      207$\pm$10  $\mu$Jy &
       $28\rlap.^s764$$\pm$$0\rlap.^{s}003$  & 04$\rlap.{''}$68$\pm$0$\rlap.{''}$05 \\   
               \bottomrule
%      \tabnotetext{a}{\scriptsize In $\mu$Jy. } 
       \tabnotetext{a}{\scriptsize For the epoch of the observations. }     
       \tabnotetext{b}{\scriptsize $23^h 07^m$.} 
      \tabnotetext{c}{\scriptsize $+21^\circ 08'$.}
    \end{tabular}
  \end{changemargin}
\end{table*}

\begin{figure*}[!t]
  \includegraphics[angle=-90,width=1.0\linewidth]{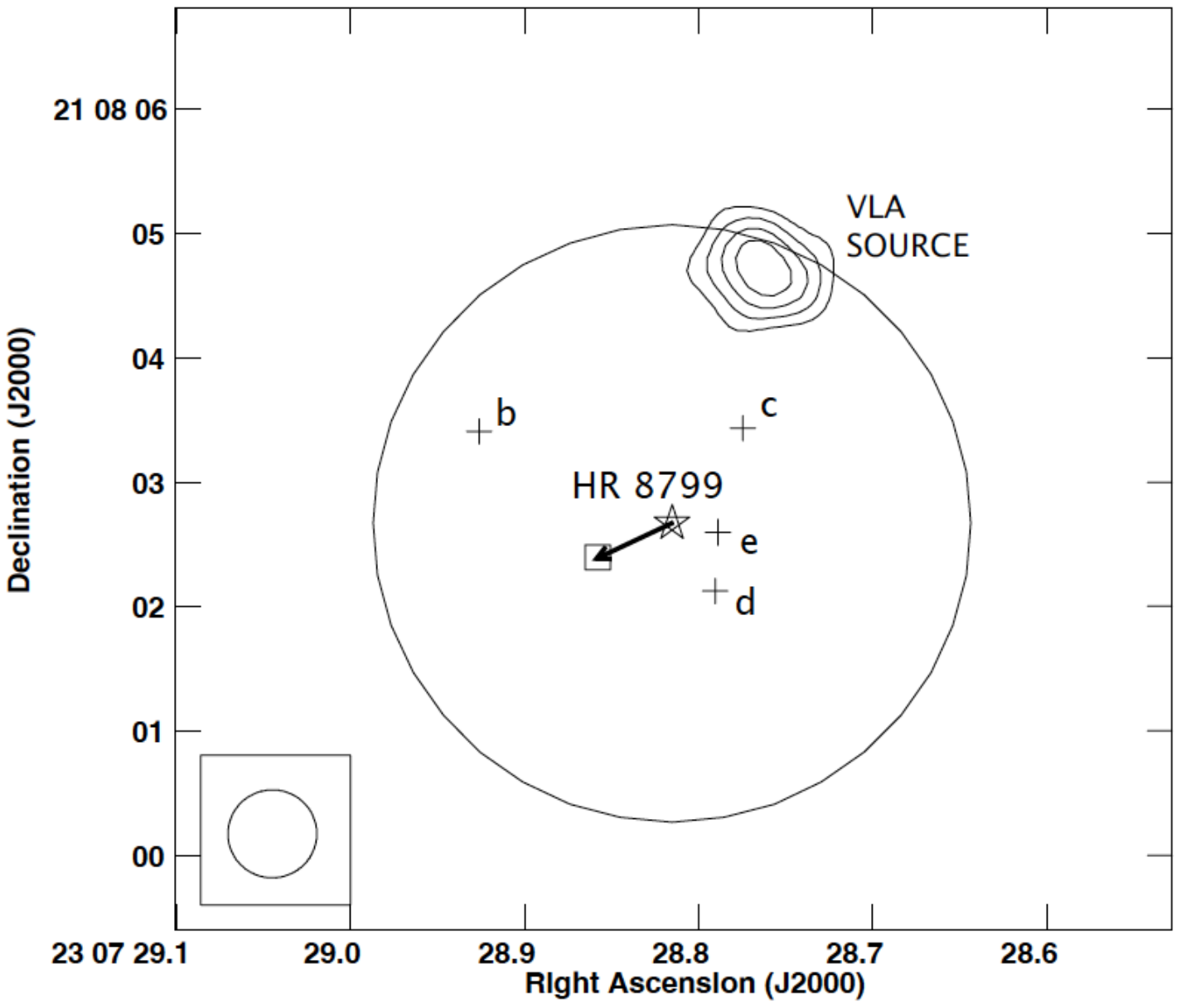}%
% \vskip-0.8cm
    \caption{\scriptsize Objects in the vicinity of HR 8799. The contours show the Very Large Array source
    observed at 3.0 GHz for epoch 2012.88.  Contours 
    are -4,  4, 6, 8 and 10
    times 1.28 $\mu$Jy beam$^{-1}$, the rms noise in this region of the image. The synthesized beam 
    ($0 \rlap.{"}71 \times 0\rlap.{"}71; -85\rlap.^\circ6$)
    is shown in the bottom left corner of the image. 
    The positions
    of HR 8799 (star) and its four exoplanets b, c, d, and e (crosses) have been
    corrected for proper motions and the orbital motions of the exoplanets to the epoch of the VLA observations (2012.88).
    The large circle marks the inner edge of the debris disk and shows that all four exoplanets as wel as the VLA source fall inside
    this inner edge. The arrow gives the proper motions of HR 8799 between
    the epoch of the VLA observations (star) and that of the ALMA observations (2018.38), with the position of HR 8799 
    for this epoch indicated with a square.
    }
  \label{fig:star}
\end{figure*}

\section{Interpretation}
\label{sec:interpretation}

\subsection{VLA data}

The final image (Figure 1) reveals the presence of a 3.0 GHz source about $2\rlap.{''}2$ to the north of HR 8799.
This radio source does not coincide in position with the star or with any of its exoplanets and appears
to be located at the inner edge of the debris disk of HR 8799. In Figure 1 we show the radio continuum source as well
as the positions of HR 8799 and its four exoplanets for epoch 2012.88, corrected 
for the proper motions of the star from Gaia DR3 
(Gaia collaboration et al. 2023) and an interpolation of the orbital motions of the exoplanets with respect to the star taken from
Konopacky et al. (2016). In Figure 1 we also show 
the approximate position of the inner radius of the debris disk. The best-fit model for this parameter given
by Faramaz et al. (2021) is 135$\pm$4 au, that at the distance of 40.85 pc derived from the Gaia DR3 parallax
equals 3.3 +- 0.1.

The parameters of the radio source are given in Table 1. No circular polarization was detected 
at an absolute 
upper limit of 19\%.
What is the \sl a priori \rm probability of finding a 3.0 GHz source with a flux density of 19.1 $\mu$Jy in a circle with radius of
$2\rlap.{''}2$? To estimate this probability we have used the 3.0 GHz source catalog of Smol{\v{c}}i{\'c} et al. (2017) to produce
a plot of the expected number of background sources as a function of flux density (Figure 2). From this figure we find that 
the expected number of sources with a flux density equal or larger than 19.1 $\mu$Jy is 1.1 per square arcmin. Since a circle
with radius of $2\rlap.{''}2$ has a solid angle of 0.0042 square arcmin, the \sl a priori \rm probability is 0.0046, or 1 in 215.
This probability is small, but not stringently improbable.

Another characteristic that may help understand the nature of the radio source is its spectral index. Dividing the data in two windows of
1 GHz wide each centered at 2.5 and 3.5 GHz, we obtain flux densities of 16.5$\pm$4.1 and 25.3$\pm$2.9 $\mu$Jy, respectively.
These flux densities give a spectral index of $\alpha$ = 1.3$\pm$0.9 ($S_\nu \propto \nu^{\alpha}$). The uncertainty is large but the value certainly favors
a positive spectral index. This is somewhat unusual for 3.0 GHz background sources since only $\simeq$11\% of them have 
spectral indices $\geq$0.4 (Smol{\v{c}}i{\'c} et al. 2017), the 1-$\sigma$ lower limit of our estimate. Finally, we searched for time variability determining the flux density of the source for each of the individual six epochs. We found that all 
individual six flux densities coincided
to the 1-2$\sigma$ level with the average value given in Table 1, suggesting no variability in the timescale of days to one month.

\begin{figure*}[!t]
  \includegraphics[width=1.0\linewidth]{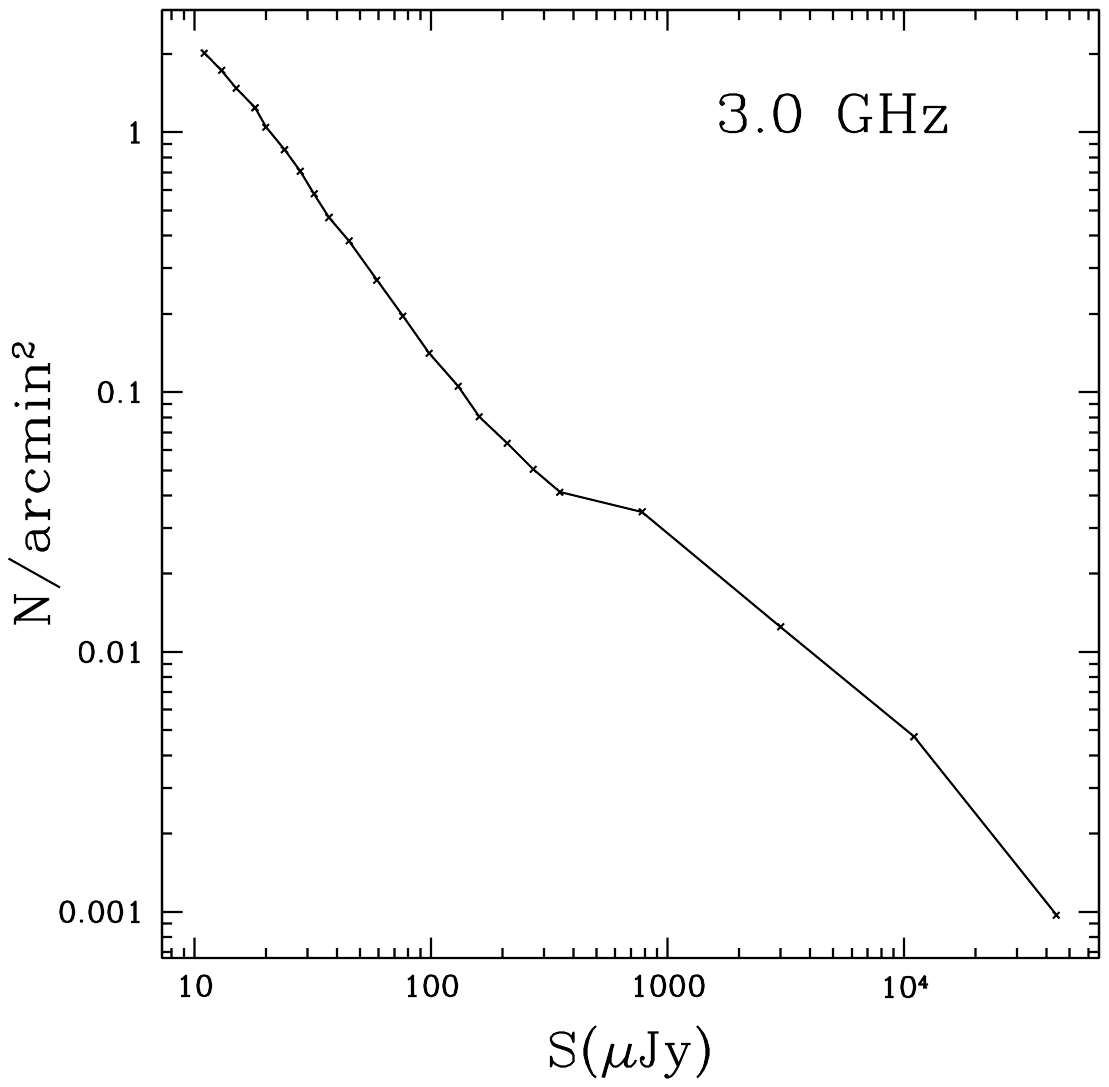}%
  \vskip-2.8cm
    \caption{\scriptsize Expected number of 3.0 GHz sources per square arcmin as function of the flux density lower limit.
    Derived from Smol{\v{c}}i{\'c} et al. (2017).
    }
  \label{fig:count}
\end{figure*}

%$RA(J2000) = 12^h ~56^m ~01\rlap.^s7927 \pm 00\rlap.^s0007; DEC(J2000) = -12^\circ ~57' ~ 25\rlap.{''}4148 \pm 00\rlap.{''}0094$. 

\subsection{ALMA data}

In the same way that at 3.0 GHz, we can ask
how probable it is that a background source with 0.49 mJy at
338 GHz falls inside the circle with a radius of  $2\rlap.{''}2$  centered on
HR 8799. From the results of Zavala et al. (2017) we estimate that the number
of background sources with a flux density of 0.49 mJy or higher
at 850 $\mu$m is about 4.9 per square arcmin. Then, the \sl a priori \rm probability 
that such a source falls inside the circle with a solid angle of 0.0042 square arcmin
is 0.021. Considering that a large number of protoplanetary disks is known, this apparent association
is not unexpected. Just in Orion, several hundred protoplanetary disks have been detected (van Terwisga et al. 2022).
As in the case of the VLA data, we searched for time variability determining the flux density of the source for each of the individual six epochs. We found that all individual six flux densities coincided
to the 1-2$\sigma$ level with the average value given in Table 1, suggesting no variability in the timescale of days to one month.

\subsection{Comparison between the VLA and the ALMA data}

Our first conclusion is that despite the low probability of finding a centimeter source so close to HR 8799, the coincidence of
positions between the VLA and ALMA data (separated by 5.5 years) rules out the association 
of the radio source with the HR 8799 system. The two positions coincide within $\leq 0\rlap.{''}03$.
HR 8799 has large proper motions, of order $0\rlap.{''}1$ per year 
($\mu_\alpha ~cos \delta = 108.284\pm0.056~mas~yr^{-1}; \mu_\delta = -50.040\pm0.059~mas yr{-1}$; 
Gaia collaboration et al. 2023). In the time interval between the two data sets, we expect
a total displacement of $\simeq 0\rlap.{''}6$, twenty times larger than our upper limit. This rules out a possible true association 
between the radio source and the HR 8799 system.

The deconvolved dimensions of the VLA and ALMA sources are consistent within the noise.
The spectral index between the 3.0 and 340.0 GHz flux densities is 0.69$\pm$0.03. This spectral index and the lack of proper motions
suggest that the radio source is a background millimeter galaxy.

We also note that the data points of Faramaz et al. (2021) have flux densities
of  316$\pm$20 and 58$\pm$18 $\mu$Jy, at 340 and 230 GHz, respectively. These two points give a spectral index of 
4.0$\pm$0.9. As noted by Faramaz et al. (2021), this value is too steep to be consistent with the typical millimeter spectral index expected from debris disks (2.5$\pm$0.4; MacGregor et al. 2016) and is more consistent with typical values expected for extragalactic dust emission (3.6$\pm$0.4; Casey 2012). This source is thus likely to be a background millimeter galaxy. Our 3 GHz flux density 
(19.1$\pm$2.7 $\mu$Jy) far exceeds the extrapolation of the mm fit to 3 GHz, and at this frequency has a spectral index of 1.3$\pm$0.9, suggesting this emission is partially thick synchrotron emission from the galaxy. Recent reviews on protoplanetary disks and debris disks 
are given by Andrews (2020) and Hughes et al. (2018), respectively.

\subsection{The Arp Effect}

These unlikely associations between sources in the plane of the sky can be called the Arp Effect.
Alton Arp (1927-2013) was a distinguished astronomer that studied interacting and apparently interacting galaxies.
In some cases, he found that two apparently interacting galaxies close in the sky had different redshifts, leading him to question the
cosmological interpretation of redshifts. Several of these apparently interacting sources are described in his Atlas of Peculiar Galaxies (Arp 1966). An example of this type of sources is Stephan's Quintet, that more properly should be called Stephan's Quartet because one of the galaxies is much closer than the other four. Arp's Effect can be summarized as follows: if you study a sufficiently large number of sources, you will find some with unlikely apparent associations. Nevertheless, the topic deserves more research.

\section{Conclusions}

We analyzed VLA observations of the A5/F0 V star HR 8799, detecting a source at $2\rlap.{''}2$ to the north of HR 8799.
The \sl a priori \rm probability of finding such a source is 0.0042. Despite this low probability that suggests a
true association, the lack of proper motions obtained when comparing the VLA data with ALMA data taken 5.5
years after favors
the radio source being a background millimeter galaxy unrelated to the HR 8799 system.

As more protoplanetary and debris disks are studied, other apparent associations are expected to emerge. 
Another case is HD 95086, where a millimeter source was found projected on its debris disk (Zapata et al. 2018).
In this case the lack of proper motions also favors that the source is a  background millimeter galaxy.
Another example is the source dubbed the "Great
Dust Cloud"  (G{\'a}sp{\'a}r et al. 2023) apparently associated with Fomalhaut, that has also been interpreted as a background millimeter galaxy
(Kennedy et al. 2023).

\begin{acknowledgments}
%\section{Acknowledgements}

This research has made use of the NASA Exoplanet Archive, which is operated by the California 
Institute of Technology, under contract with the National Aeronautics and Space Administration under the Exoplanet Exploration Program. This research has made use of data obtained from or tools provided by the portal exoplanet.eu of The Extrasolar
Planets Encyclopaedia. This work also made use of data from the European Space Agency (ESA) mission
{\it Gaia} (\url{https://www.cosmos.esa.int/gaia}), processed by the {\it Gaia}
Data Processing and Analysis Consortium (DPAC,
\url{https://www.cosmos.esa.int/web/gaia/dpac/consortium}). Funding for the DPAC
has been provided by national institutions, in particular the institutions
participating in the {\it Gaia} Multilateral Agreement.  
L.A.Z. acknowledges financial support from CONACyT-280775 and UNAM-PAPIIT IN110618, 
and IN112323 grants, M\'exico. 
L.F.R. acknowledges the financial support of DGAPA (UNAM) 
IN105617, IN101418, IN110618 and IN112417 and CONACyT 238631 and 280775-CF grant 263356.

\end{acknowledgments}

\end{document}